\documentclass[useAMS,usenatbib]{mn2e}
\usepackage{graphicx}
\usepackage{graphics}
\usepackage{epstopdf}
\usepackage{float}
\usepackage{bm}
\usepackage{epsf}
\usepackage{url}

\title[The Hierarchical Nature of the  Spin Alignment of Dark Matter Haloes in Filaments]
{The Hierarchical Nature of the  Spin Alignment of Dark Matter Haloes in Filaments}
\author[Aragon-Calvo M.A.]{M.A. Aragon-Calvo$^{1}$\thanks{E-mail:miguel@pha.jhu.edu} \\
$^{1}$The Johns Hopkins University, 3400 Charles St., Baltimore, MD, USA.\\}
\begin{document}

%\date{Submitted to MNRAS}

\pagerange{\pageref{firstpage}--\pageref{lastpage}} \pubyear{2002}
\maketitle
\label{firstpage}

%========================================================================
% ABSTRACT
%========================================================================
\begin{abstract}

Dark matter haloes in cosmological filaments and walls have their spin vector aligned (in average) with their host structure.
While haloes in walls are aligned with the plane of the wall independently of their mass, haloes in
filaments present a mass dependent two-regime orientation.
Here we show that the transition mass determining the change in the alignment regime (from parallel to perpendicular)
depends on the hierarchical level in which the halo is located, reflecting the hierarchical nature of
the Cosmic Web. By explicitly exposing this hierarchy we are able to identify
the contributions of different components of the filament network to the spin alignment signal.
We discuss a unifying picture to describe the alignment of haloes in filaments and walls consistent
with previous results and our findings based on a two-phase angular momentum acquisition, first
via tidal torquening and later via anisotropic mass accretion.

The hierarchical identification and characterization of cosmic structures was done with a new implementation
of the Multiscale Morphology Filter. 
The MMF-2 (introduced here) represents a significant improvement over its predecessor in terms of robustness 
and by explicitly describing the hierarchical relations between the elements of the Cosmic Web.

\end{abstract}

%==========================================================================
%--- Introduction
%==========================================================================
\section{Introduction}

In the gravitational instability scenario galaxies acquire their angular momentum via
tidal torque produced by a misalignment between the inertia tensor of the proto-halo and the
embedding gravitational shear tensor generated by the surrounding matter distribution. This is
the so called Tidal Torque Theory (TTT) \citep{Hoyle51,White84}. Although the TTT predicts
a correlation between the spin orientation of galaxies and their local Large Scale Structure
observations of spin alignments have been so far inconclusive and often contradictory
(\citet{Kashikawa92,Navarro04,Patiri06,Trujillo06,Lee07,Brunino07,Jones10,Varela12,Tempel13} among others).
The reason is still not clear and may include the intrinsic difficulties in deducing the spin vector 
of galaxies, small galaxy samples, challenges in characterizing LSS from galaxy redshift 
catalogues, etc. 

On the theoretical/numerical front the picture is more clear. To date there is a consensus that the 
spin vector of haloes in walls lies in the plane of their host walls while haloes in filaments have their spin
either parallel or perpendicular to their host filament depending on their mass \citep{Aragon07a,Hahn07,Zhang09,Codis12,Trowland13,Libeskind13}.
An important aspect of the spin alignment of haloes in filaments is the existence of a transition mass M$_{\textrm{\tiny{tr}}}$, at which 
haloes change their orientation from parallel to perpendicular. This effect was first reported by \citet{Aragon07a}
and later confirmed by other authors \citep{Hahn07,Zhang09,Codis12,Trowland13,Libeskind13}.
The spin alignment of low-mass haloes also has a redshift dependence starting perpendicular and becoming parallel
at $z \sim 1$ \citep{Aragon07c,Trowland13}. This common primordial orientation and subsequent change indicates an
additional mechanism for angular momentum acquisition that has a stronger effect in low-mass haloes.
In the observational front the two regimes of alignment/anti-alignment of low-mass and high-mass haloes in filaments have been 
recently observed by \citet{Tempel13} if one assumes a correlation between Hubble type and galaxy/halo mass. They found that bright spirals 
have a weak tendency to have their spin parallel to their host filaments while Elliptical/S0 show a strong perpendicular alignment.

The existence of a hierarchy of structures in the Cosmic Web in which large structures contain and are defined by smaller ones 
(\citet{Sheth04, Einasto11, Aragon13} among others) begs the question of whether the halo alignment
reflects this hierarchy. The limited number of studies that link the alignment signal and the hierarchy of cosmic structures 
indicate that the transition mass of alignment in filaments depends on the scale of the filament \citep{Codis12}
and the distance from the halo to the closest cluster \citep{Zhang09}. However, to date no direct link has been
stablished between halo alignment and the hierarchy of cosmic structures.

The TTT theory does not provide a mechanism (at least a simple one) to produce the observed two-regime alignment of haloes with their mass or to 
change the alignment of haloes from perpendicular to parallel with time.
Recently an alternative mechanism for angular momentum acquisition has been discussed by \citet{Libeskind12,Codis12} and \citet{Trowland13}
in which angular momentum is generated by the anisotropic infall of matter. This is in term defined by the
surrounding LSS morphology and dynamics, offering a natural process for spin-LSS correlations. This secondary-torque
mechanism may help explain the change in alignment and possibly even contribute to the primordial torque generation.

%-------------------
%----  FIGURE  -----
%-------------------
\begin{figure*}
  \centering
  \includegraphics[width=0.99\textwidth,angle=0.0]{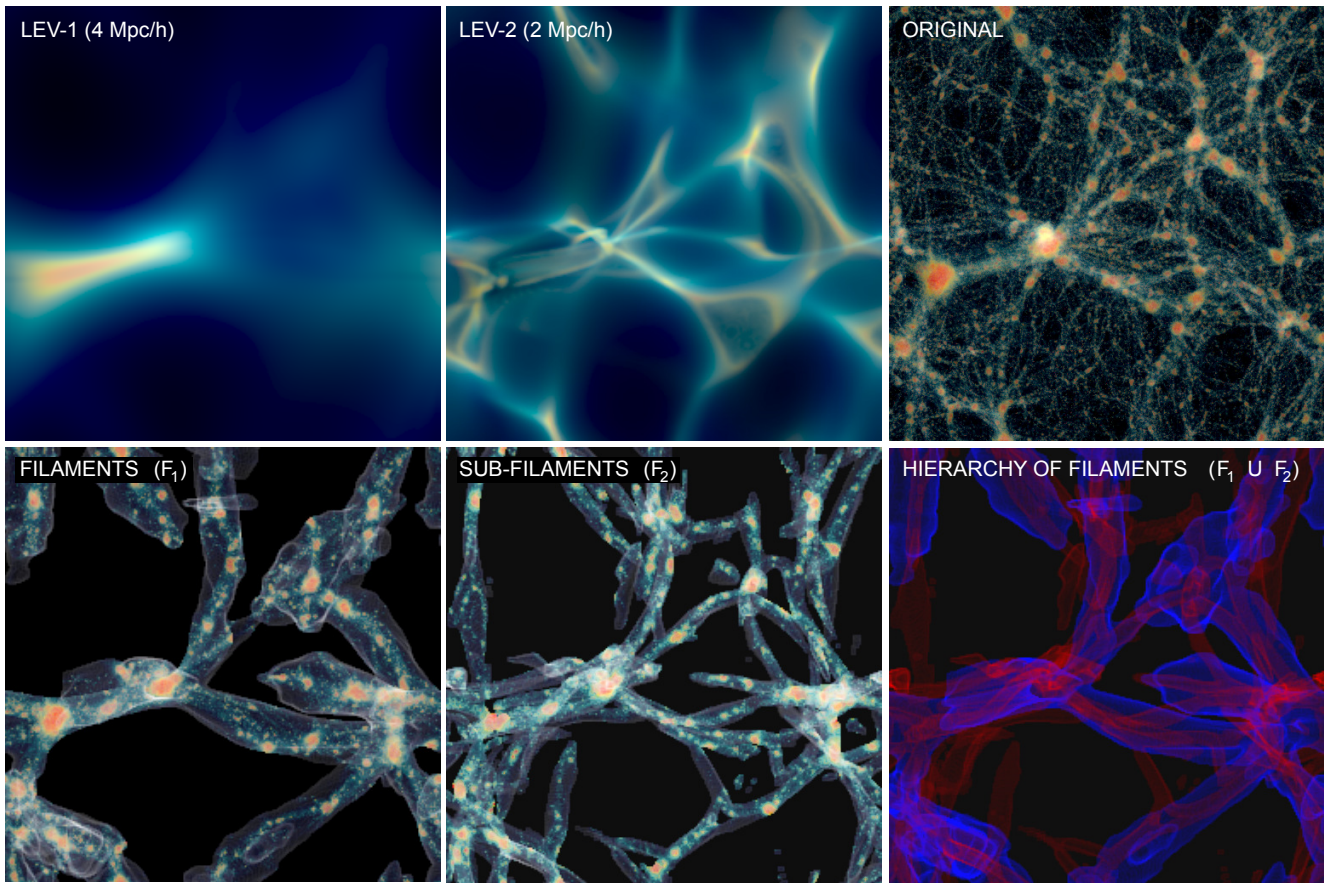}
  \caption{The MMF-2 method. Top panels: hierarchical space. We show the 
			volume rendering of the density field at $z=0$ for linear-regime smoothing at 4, 2 Mpc h$^{-1}$ 
			and no smoothing (left, center and rights panels respectively).
			The thickness of the slices was chosen to show as many structures as possible while avoiding confusion.			Bottom panels: Hierarchical identification of filaments. The left and center panels show the density field inside filaments (F$_1$)
			and sub-filaments (F$_2$) respectively. We highlight the filament mask as a semitransparent surface. The hierarchy
			of filaments (blue) and sub-filaments (red) is shown in the bottom right panel.}
  \label{fig:density_hierarchical}
\end{figure*}

%==========================================================================
%   MMF 2.0
%========================================================================== 
\section{MMF-2: Hierarchical Identification of Filaments and Walls}\label{sec:MMF}

In order to study the effect of environment in the alignment of haloes in filaments and walls we need
to characterize the density field according to its local geometry, compute its local direction and,
for the problem at hand, its \textit{hierarchical relations}. 
The identification of filaments and walls was performed using a significantly improved implementation of the 
Multiscale Morphology Filter (MMF, \citet{Aragon07b}). The MMF-2 performs the identification of structures on a generalization 
of scale-space: the hierarchical-space, which exposes the \textit{hierarchical character} of the density field in contrast to the
scale-space approach (as in the original MMF and its derived implementations \citep{Zhang09,Cautun13})
which emphasize the \textit{scale} of the structures and is insensitive to their nesting relations.
The hierarchical space used in the MMF describes both scale and nesting relations.

We generate the hierarchical space as described in  \citet{Aragon10b} and \citet{Aragon12}, by Gaussian-smoothing 
the \textit{initial conditions} instead of the \textit{final} density field as in the scale-space approach.
This \textit{linear-regime smoothing} is applied when all Fourier modes are independent, allowing us to target
specific scales in the density field before Fourier mode-mixing occurs. The smooth initial conditions evolve by gravity 
producing all the anisotropic features of the Cosmic Web but lacking of small-scale structures including dense
haloes (below the smoothing scale). This reduces the dynamic range in the density field and greatly limits
the contamination produced by dense haloes in the identification of filaments and walls. 

%The hierarchical space is analog to the \textit{deep structure} analysis introduced by \citet{Koenderink84}.
The identification of structures in the original MMF method and the one presented here is done based on the second-order local variations 
of the density field encoded in the Hessian matrix. LSS morphologies are associated to ratios between the 
eigenvalues of the Hessian matrix ($\mathbf{\lambda_1} < \mathbf{\lambda_2} < \mathbf{\lambda_3}$, see \citet{Aragon07b}). 
The output of the MMF-2 is a set of ``masks" sampled on a regular grid indicating which voxels
belong to a given morphology at a given hierarchical level. For our purposes we generated a filament, sub-filament
wall and sub-wall masks (F$_1$, F$_2$, W$_1$ and W$_2$ respectively, the subscript indicates the hierarchical level). These masks 
roughly correspond to the ``feature maps" $\mathcal{F}$ in the original MMF algorithm. 
A complete description of the MMF-2 method will be presented in Aragon-Calvo (2013) in preparation.

%==========================================================================
%   SIMULATION
%==========================================================================
\section{N-body Simulation and DM haloes}

The N-body simulations and related halo catalogues used in this work are based on the MIP
correlated ensemble simulation. \citep{Aragon12}. The MIP simulation consists of 220 realizations 
sharing identical LSS fluctuations defined at $k < k_{\textrm{\tiny{cut}}}$ where $k_{\textrm{\tiny{cut}}}$ corresponds to a scale 
of 4 Mpc h$^{-1}$. At smaller scales ($k > k_{\textrm{\tiny{cut}}}$) realizations can be considered independent
between each other. 
Each realization in the ensemble contains $256^3$ dark matter particles inside a box of 32 Mpc h$^{-1}$ of side
with cosmological parameters $\Omega_m = 0.3, \Omega_{\Lambda} = 0.7, h = 0.73, \sigma_8 = 0.8$.

%----------------------
%
%----------------------
\subsection{Hierarchical space}
For the purpose of this paper we generated a two-level hierarchical space defined by linear-regime smoothing
scales of $4$ and $2$ Mpc h$^{-1}$ delineating the prominent structures in the Cosmic Web
and their substructure respectively. We denote these hierarchical levels as  LEV-1 (filaments/walls) and LEV-2 (sub-filaments/sub-walls). 
From the template initial conditions used to generate the MIP ensemble (see \citet{Aragon12} for details) 
we generated low-resolution $128^3$ initial conditions grids and smoothed them at the scales defined above. 
We then evolved the particles to the present time $z=0$ using the PM-Tree code {\small GADGET-2} \citep{Springel05}.
These simulation were used only to compute density fields.

%----------------------
%
%----------------------
\subsection{Density fields}
From the particle distribution we computed densities following the technique presented in \citet{Abel12} and \citet{Shandarin12}. In this 
novel approach particles define Lagrangian volumes that reflect the changes in density as particles evolve
deforming the original volume. Lagrangian volumes are defined by 8 adjacent particles forming a ``cube" that is divided into
6 tetrahedra. We estimate densities in the tessellation as the mean density of the adjacent tetrahedra to
each vertex and interpolated to a regular grid taking care to account for multistreams.
This method makes density estimation computationally straightforward,
it produces a continuous volume-filling density field with practically no artifacts in low and medium-density regions 
(our regions of interest) and is self-adaptive by construction.

%-------------------
%----  FIGURE  -----
%-------------------
\begin{figure*}
  \centering
  \includegraphics[width=0.99\textwidth,angle=0.0]{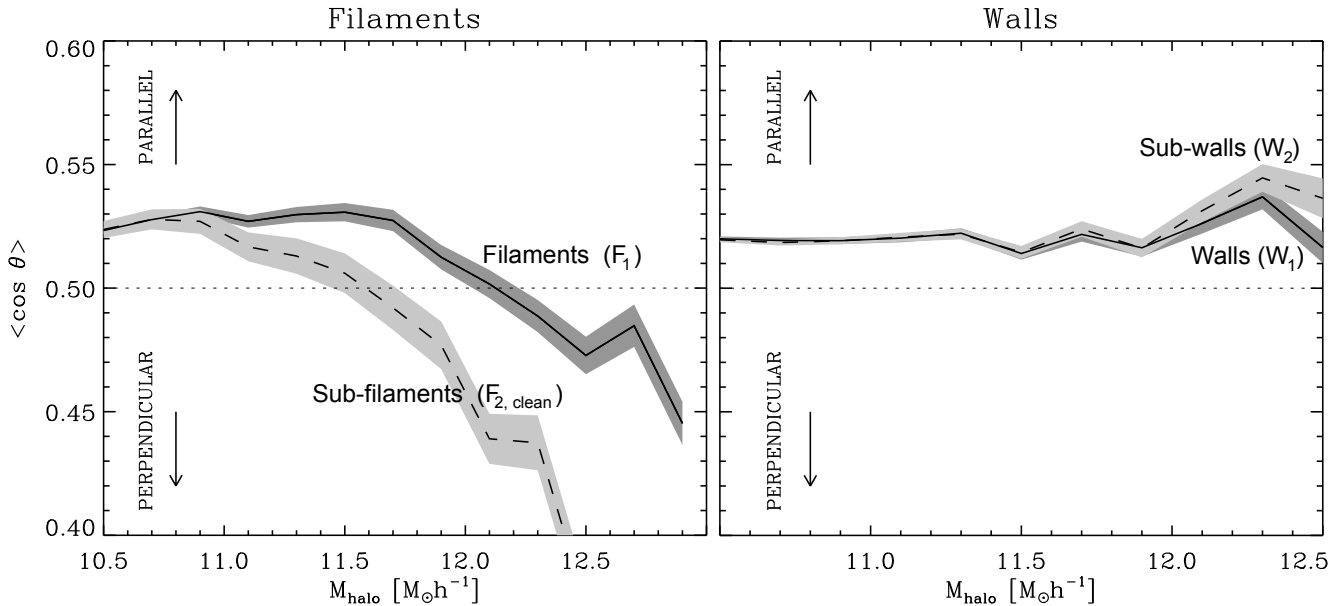}
  \caption{Spin alignment of haloes as function of mass in filaments (left panel) and walls (right panel). The alignment is measured
		with the mean $\cos \theta $ where 0.5 corresponds to no preferred alignment. The sample was divided in
		haloes in filaments/walls (solid lines) and sub-filaments/sub-walls (dashed lines). 
		Error bars per bin were computed from 1000 random realizations with the same number of points as their corresponding bin.}
  \label{fig:mean_cos_theta}
\end{figure*} 

%----------------------
%
%----------------------
\subsection{Halos}

Haloes were identified in each realization of the MIP ensemble using the friends-of-friends algorithm with a linking parameter $b=0.2$. 
We imposed a count limit of 50 particles corresponding to a halo mass of  $\sim 8 \times10^{9}$ M$_{\odot}$ h$^{-1}$.
We use the MIP ensemble in ``stack mode", where halo catalogues from all realizations are aggregated into one ``master halo catalogue". The full 
MIP stacked ensemble contains $\sim 5\times 10^6$ haloes inside a volume of (32 Mpc h$^{-1}$)$^3$ giving unprecedented halo number density. 
For each FoF group, we compute its mass, center of mass, mean velocity etc. The angular momentum was computed as:

\begin{equation}
\mathbf{J} = \sum_i^N m_i \;\; \mathbf{r_i} \times (\mathbf{v_i} - \mathbf{\bar{v}}),
\end{equation}

\noindent where the sum is over all the particles in the halo, $m_i$ is the particle mass, $\mathbf{r_i}$ is the distance 
to each particle from the center of the halo, $\mathbf{v_i}$ is the peculiar velocity of the particle and 
$\mathbf{\bar{v}}$ the mean velocity of the halo with respect to its center of mass. 

%----------------------
%
%----------------------
\subsection{Spin alignment in filaments and walls}
The angle between the spin of the halo and its host filament or wall was computed as
$\theta_F = \phi_W $ and $\theta_W = 90^{\circ} - \phi_W $, where 
$\phi_F = \mid \cos^{-1} (\mathbf{J} \cdot \bf{\lambda_1}) \mid $ and 
$\phi_W = \mid \cos^{-1} (\mathbf{J} \cdot \bf{\lambda_3}) \mid $ are the angles between the spin of 
the halo and the direction of its host filament ($\mathbf{\lambda_1}$ corresponding to the direction of constant density along the 
filament's axis) or the normal to the wall ($\mathbf{\lambda_3}$ corresponding to the direction of maximum change in the density
field).

%----------------------
%
%----------------------
\subsection{The filament/sub-filament sample}
Figure \ref{fig:density_hierarchical} shows the density field at $z=0$ from the
two hierarchical levels defined by 4 and 2 Mpc h$^{-1}$ linear-regime smoothing (LEV-1 and LEV-2 respectively)
and the original initial conditions with no smoothing (top panels from left to right) . The original density field 
enclosed in the filament and sub-filament mask is shown at the bottom-left and center panels. The two-level hierarchy of filaments 
(the union of the two sets) is shown in the bottom-right panel. 
%Walls and sub-walls (not shown here) are delineated by their corresponding network of filaments or sub-filaments.

In order to clearly differentiate between the alignment signal of haloes in filaments and sub-filaments we 
``cleaned" the sub-filament mask by removing the sub-filaments embedded inside
filaments. We also constrained sub-filaments to be contained by walls. This removes the tenuous intra-void filaments. 
The mask of ``clean" sub-filaments is then defined as:

\begin{equation}
\textrm{F}_{2,\textrm{\tiny{clean}}} = \textrm{F}_2 \cap \textrm{W}_1 \cap \textrm{F}_{1}^{\textrm{\tiny{C}}}
\end{equation}

\noindent where F$_2$ is the sub-filament mask, W$_1$ is the wall mask and F$_{1}^{\textrm{\tiny{C}}}$ is the 
complement of the filament mask. In what follows F$_1$ and F$_{2,\textrm{\tiny{clean}}}$ denote the masks 
we used to compute spin-alignment of haloes in filaments and sub-filaments respectively

%==========================================================================
%--- Results
%==========================================================================
\section{Results}

\begin{figure}
  \centering
  \includegraphics[width=0.45\textwidth,angle=0.0]{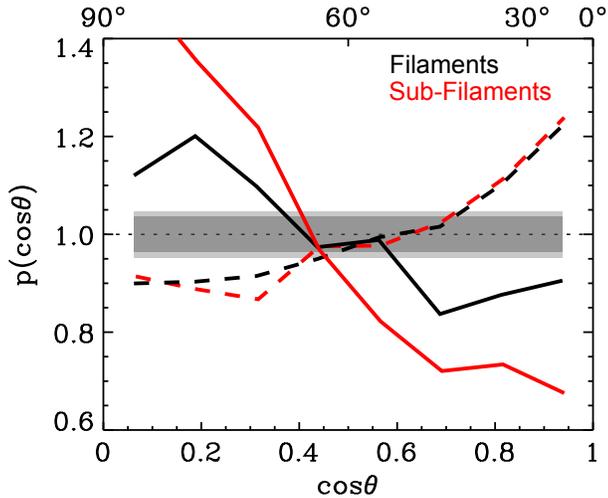}
  \caption{Distribution of angles between the spin of a halo and its parent filament (black line) or sub-filament (red line). The samples were divided
			as high-mass haloes (solid line) and low-mass haloes (dashed line). The transition mass for filaments and sub-filaments corresponds
			to $6 \times 10^{11}$ and $2 \times 10^{12}$ M$_{\odot}$ h$^{-1}$ respectively.}
  \label{fig:halo_spin_distribution}
\end{figure}

Our main findings are shown in Figure \ref{fig:mean_cos_theta} where we measure the mean cosine of the angle between the spin vector of
haloes and the direction of their host filament/sub-filament (left panel) or wall/sub-wall (right panel) as a function of halo mass. 
A more detailed view of the spin alignment distribution for haloes in filaments/sub-filaments is shown in Figure \ref{fig:halo_spin_distribution}.
Our results can be summarized as follows:

\begin{itemize}

\item Haloes in filaments and sub-filaments present a two-regime spin alignment as a function 
of their mass: low-mass haloes are (in average) aligned parallel to their parent filament while high-mass haloes are oriented perpendicular to it, 
confirming previous findings.

\item The mean alignment of low-mass haloes seems to converge at $< \cos \theta> \sim 0.53$ down to a halo mass of
$\sim 10^{10}$ M$_{\odot}$ h$^{-1}$ in both filaments and sub-filaments.

\item The alignment of high-mass haloes does not seem to converge like in the case of the low-mass haloes. Instead, 
the alignment signal increases with increasing mass.

\item We found a significant difference in the transition mass M$_{\textrm{\tiny{tr}}}$ dividing parallel and perpendicular alignment 
for haloes in filaments and sub-filaments being M$_{\textrm{\tiny{tr,low}}} \sim 1.5 \times 10^{12}$ M$_{\odot}$ h$^{-1}$ and  
M$_{\textrm{\tiny{tr,high}}} \sim 4 \times 10^{11}$ M$_{\odot}$ h$^{-1}$ respectively. 

\item The slope in the alignment-mass curve of high-mass haloes in sub-filaments is more pronounced than in the case of haloes in filaments.

\item Haloes in walls/sub-walls have their spin oriented in the plane of the wall for the full range of masses in our halo sample
($\sim 10^{10} - 10^{12.5}$ M$_{\odot}$ h$^{-1}$ ).

\item While the spin alignment distribution of haloes in filaments and sub-filaments seem scaled versions of each other 
(see Figure \ref{fig:halo_spin_distribution}) the spin alignment distribution of low-mass haloes is practically 
the same for filaments and sub-filaments even though their transition masses are different by a factor of 3.

\end{itemize}

%==========================================================================
%--- Conclusions
%==========================================================================
\section{Conclusions and discussion}

Using an advanced LSS classification algorithm (MMF-2) and a novel N-body correlated ensemble simulation
we were able to measure the halo spin alignment signal with unprecedented detail in a variety of cosmic environments.
Our results confirm previous findings while significantly improving the alignment signal detection.
However, we found a difference in the transition mass M$_{\textrm{\tiny{tr}}}$ at which the spin alignment changes from being parallel (low-mass haloes) to 
perpendicular (high-mass haloes) depending on the hierarchical level of the host filament.
This hierarchy-dependent effect has not been observed before due to the inability 
to explicitly target the nesting relations of the structures in the Cosmic Web. 

We found that haloes in walls have their spine aligned with the plane of the wall independently of the halo's mass.

The explicit link between halo spin alignment and Cosmic Web hierarchy was possible with the use of the 
MMF-2 method (introduced here) that identifies cosmic structures and their hierarchical
relations and benefits from an improved density estimation algorithm.
The use of the MIP simulation allowed us to perform a high-resolution analysis of the density field over a relatively
small volume while at the same time providing us with the large number of haloes needed to detect the spin alignment signal.

%----------------------
%
%----------------------
\subsection{On the origin and fate of the spin alignment}

The change of alignment of haloes in filaments, the dependence of the transition mass with redshift \citep{Codis12} and 
filament hierarchy (this work) indicates a ``secondary torque" process that acts cumulatively and affects 
all haloes but manifest itself first in low-mass haloes. 
%The convergence of the alignment signal of low-mass haloes in filaments and sub-filaments is puzzling and it 
%may give clues on the nature of this ``secondary torquening" process.
Since tidal torque is mostly efficient before turnaround the process changing the spin orientation of haloes must
be of a different nature. The anisotropic infall scenario offers a tantalizing alternative. Walls
and sub-walls accrete mass from their adjacent voids in the direction parallel to the normal of the wall
inducing angular momentum on the plane of the wall so haloes retain their primordial spin alignment.
On the other hand filaments and sub-filaments accrete mass mainly from their adjacent walls in the
direction perpendicular to the filament thus inducing a parallel spin alignment. Low mass haloes are more
affected because their low-mass makes then prone to have their spin direction changed for a given 
mass accretion rate compared to high-mass haloes.
The stronger perpendicular alignment of haloes in sub-filaments in comparison to haloes in filaments reflects 
the more quiet environment characteristic of walls and their embedded sub-filaments. This is a natural
consequence of the hierarchy of the Cosmic Web.

%where they fitted a function 
%$M_t = M_0 (1+z)^{-\gamma}$ where $M_0$ was similar to the value first found by \citet{Aragon07}.

%----------------------
%
%----------------------
%\subsection{MMF-2 Hierarchical identification of structures}

%The method presented here can be applied to N-body simulations with relatively little computational overhead since
%the creation of the hierarchical space can be done with a lower number of particles than the ``main" simulation
%as presented in this work. 

%The hierarchical identification of structures as presented here is limited to numerical simulations where one can
%perform the linear-regime smoothing. For applications such as spectroscopic galaxy catalogues the most viable
%option is to generate a scale-space with a smoothing function such as the Gaussian. In this case the method
%falls back to the original MMF algorithm.

%-------------------
%----  FIGURE  -----
%-------------------

%==========================================================================
%--- Acknowledgements
%==========================================================================
\section{Acknowledgements}
This research was funded by the Betty and Gordon Moore Foundation and 
a ``New Frontiers of Astronomy and Cosmology" grant from the Sir John Templeton Foundation.
The author would like to thank Mark Neyrinck, Xin Wang, Lin Yang and Alex Szalay for stimulating discussions.

\end{document}